\documentclass[twocolumn,showpacs,showkeys,pre]{revtex4}

\usepackage{dcolumn}
\usepackage{amsmath}
\usepackage{amssymb}

\usepackage{graphicx}
\usepackage{graphics}

\begin{document}

\title{Network properties of protein structures}

\author{Ganesh Bagler}
\email{ganesh@ccmb.res.in}
\author{Somdatta Sinha}
\email{sinha@ccmb.res.in}
\affiliation{Centre for Cellular and Molecular Biology, Uppal Road, Hyderabad 500007, India}
\date{\today}
\begin{abstract}
Protein structures can be studied as complex networks of interacting amino acids. We study proteins of different structural
classes from the network perspective. Our results indicate that proteins, regardless of their structural class, show small-world
network property. Various network parameters offer insight into the structural organisation of proteins and provide indications 
of modularity in protein networks.
\end{abstract}
\maketitle

\section{Introduction}
\label{sec:intro} 
Biological systems have been studied as networks at different levels: protein-protein interaction
network~\cite{protein:yook}, metabolic pathways network~\cite{metabolic:jeong,ravsaz:science}, gene regulatory
network~\cite{gene:reka}, and protein as a network of amino acids~\cite{cabios,protnet:PRE,protnet:JMB,protnet:Biophys}. 
Proteins are biological macromolecules made up of a linear chain of amino acids and are organised into
three-dimensional structure comprising of different secondary structural elements. They perform diverse biochemical functions
and also provide structural basis in living cells. It is important to understand how proteins consistently fold into their
native-state structures and the relevance of structure to their function. Network analysis of protein structures is one such
attempt to understand possible relevance of various network parameters.

There have been several efforts to study proteins as networks. Asz\'{o}di and Taylor~\cite{cabios} compared the linear chain of
amino acids in a protein and its three dimensional structure with the help of two topological indices -- connectedness and effective
chain length -- related to path length and degree of foldedness of the chain. Kannan and Vishveshwara~\cite{kannan} have used the
graph spectral method to detect side-chain clusters in three-dimensional structures of proteins. In recent years, with
the elaboration of network properties in a variety of real networks, Vendruscolo et al.\/~\cite{protnet:PRE} showed that
protein structures have small-world topology. They also studied transition state ensemble (TSE) structures to identify the key
residues that play a key role of ``hubs'' in the network of interactions to stabilise the structure of the transition state.
Greene and Higman~\cite{protnet:JMB} studied the short-range and long-range interaction networks in protein structures and showed
that long-range interaction network is \emph{not} small world and its degree distribution, while having an underlying scale-free
behaviour, is dominated by an exponential term indicative of a single-scale system. Atilgan et al.\/~\cite{protnet:Biophys}
studied the network properties of the core and surface of globular protein structures, and established that, regardless of size, the
cores have the same local packing arrangements. They also explained, with an example of binding of two proteins, how the small-world 
topology could be useful in efficient and effective dissipation of energy, generated upon binding.

In this study, we model the native-state protein structure as a network made of its constituent amino-acids and their
interactions. The $C_{\alpha}$ atom of the amino acid has been used as a node and two such nodes are said to be linked if they
are less than or equal to a threshold distance apart from each other~\cite{protnet:PRE,protnet:Biophys}. We use $7$\AA\ as the
threshold distance. Our results show that proteins are small-world networks regardless of their structural classification across four
major groups as enumerated in Structural Classification of Proteins (SCOP)~\cite{SCOP}. We also highlight the differences in
some of the network properties among these classes. Our studies are indicative of the modular nature of these networks.

\begin{figure*}
\begin{center}
\begin{tabular}{cc}
\includegraphics[width=6.5cm]{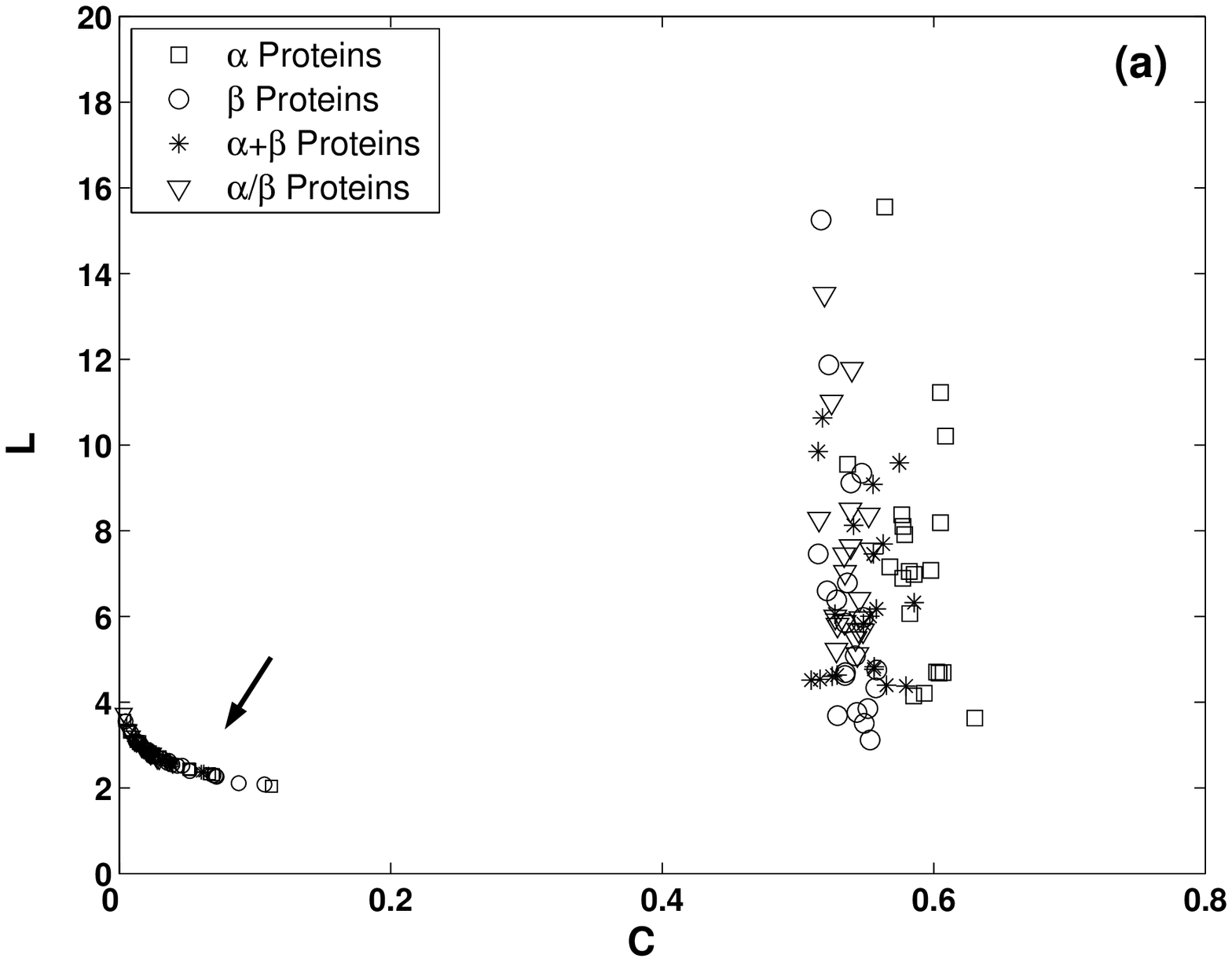} &
\includegraphics[width=6.5cm]{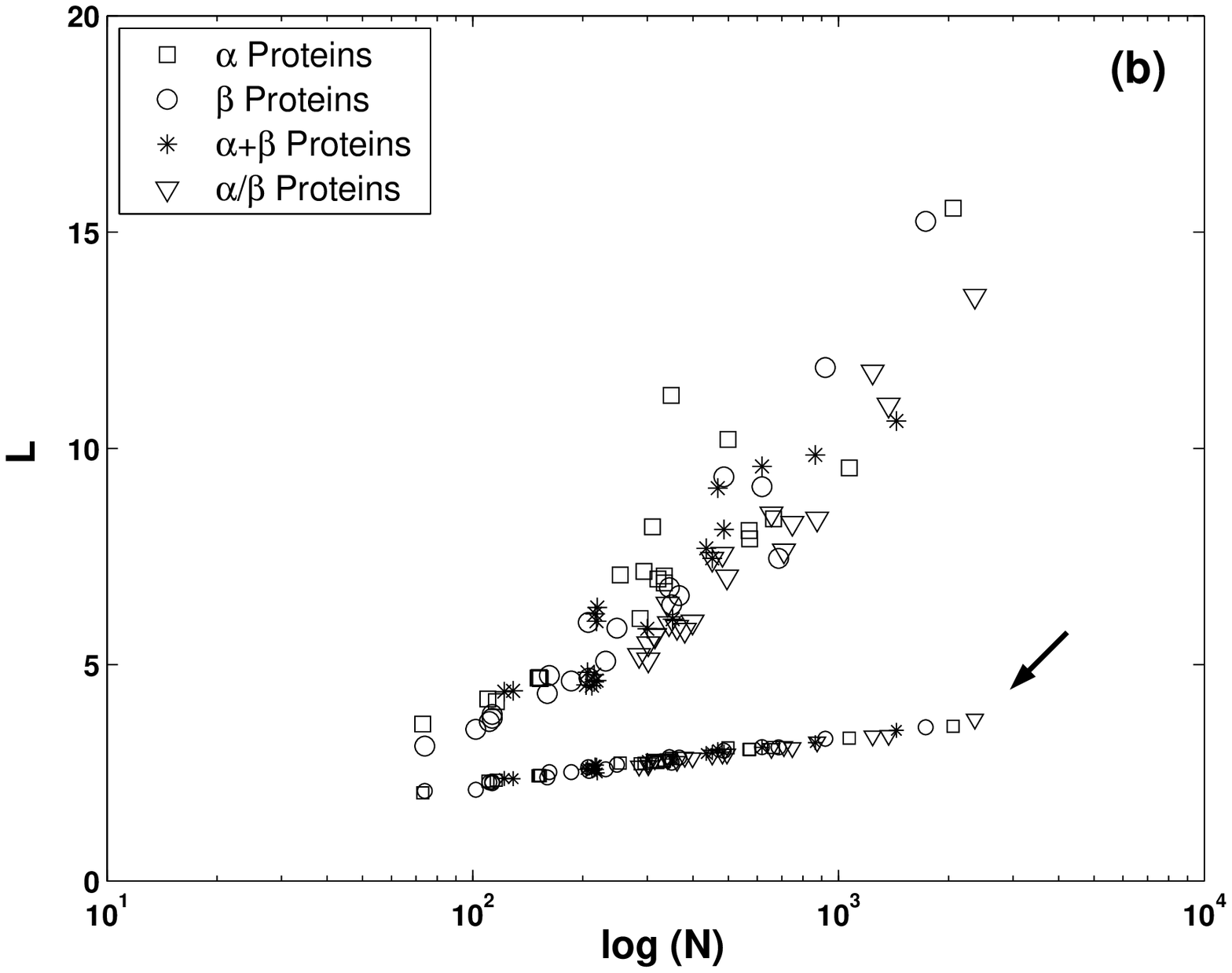}
\end{tabular}
\end{center}
\caption{(a) The L--C plot of proteins from four structural classes. (b) Increase in the $L$ of proteins with logarithmic increase in 
size ($N$). Random controls are indicated by an arrow in both the figures.}
\label{fig:fig1}
\end{figure*}

\section{Methodology}
\label{sec:methodology}
The four structural classes (from SCOP) of proteins chosen are: $\alpha$, $\beta$, $\alpha+\beta$, and
$\alpha/\beta$. The $\alpha$ proteins are composed predominantly of $\alpha$ helices, and the $\beta$ proteins of $\beta$ sheets.
The $\alpha+\beta$ proteins mainly have anti-parallel $\beta$ sheets, whereas those in $\alpha/\beta$ consist of mainly parallel
beta sheets. We consider $20$ proteins from each of these classes whose sizes range from $73$ to $2359$ amino acids. The structural
data is obtained from the Protein Data Bank (PDB)~\cite{PDB}.

The parameters used to characterise the network are :\\
(i) The \textit{Degree} ($k_i$) of a node $i$ is the number of nodes to which it is directly connected. Average degree, $K$, of a
network with $N$ nodes is defined as $$K=\frac{1}{N}{\sum_{i=1}^N k_{i}}.$$
(ii) The \textit{Average Shortest Path Length} is defined as $$L=\frac{1}{N(N-1)}{\sum_{i=1}^{N-1} \sum_{j=i+1}^N L_{ij}},$$ 
where $L_{ij}$ is the shortest path length between nodes $i$ and $j$.\\
The \textit{Diameter} ($D$) of the network is the largest of all the shortest path lengths.\\

(iii) The \textit{Clustering Coefficient} ($C_{i}$) for a node $i$ is defined as the fraction of links that exist among its nearest
neighbours to the maximum number of possible links among them. The \textit{Average Clustering Coefficient} ($C$) of the network is 
defined as $$C={\frac{1}{N}\sum_{i=1}^N C_{i}}.$$

A network is a ``small-world network'' if it has high C and if its L scales logarithmically with N~\cite{watts:book}. A network
lacking a characteristic degree and having degree distribution of a power-law form is known as 
``scale-free network''~\cite{SFNW:barabasi}.

\section{Results}
\label{sec:results} 

\subsection{Network parameters for different structural classes of proteins}
We calculate $L$ and $C$ for each protein. As controls, we calculate the $L$ and $C$ of random graphs and one-dimensional
(1-d) regular graphs of the same $N$ and $K$. Figure~\ref{fig:fig1}(a) shows the L--C plot for the proteins and
their random controls (indicated by an arrow). The averages of the distribution of $L$ and $C$ for the protein networks 
are $6.88\pm2.61$ and
$0.553\pm0.027$, respectively. The $L_{random}$ and $C_{random}$ are $2.791\pm0.348$ and $0.031\pm0.022$, and that for the
comparable 1-d regular graphs are $L_{regular}=29.0\pm25.97$ and $C_{regular}=0.643\pm0.004$. The Kolmogorov-Smirnov
test~\cite{nr:book} shows that the differences between $L$ and $C$ of the proteins and random or regular lattices (not shown in
Fig.\/~\ref{fig:fig1}(a)) are statistically significant. Thus, these protein networks have significantly high clustering
coefficient than their random counterparts and the $L$ and $C$-values fall between the random and regular networks in the
L--C plot.

%\begin{figure*}
%\begin{center}
%\begin{tabular}{cc}
%\includegraphics[width=6.5cm]{G_Bagler_fig1a.eps} &
%\includegraphics[width=6.5cm]{G_Bagler_fig1b.eps}
%\end{tabular}
%\end{center}
%\caption{(a) The L--C plot of proteins from four structural classes. (b) Increase in the $L$ of proteins with logarithmic increase in 
%size ($N$). Random controls are indicated by an arrow in both the figures.}
%\label{fig:fig1}
%\end{figure*}

Fig.\/~\ref{fig:fig1}(b) shows $L$ of all proteins with different $N$ and their random counterparts (indicated by an arrow). 
It can be seen that $L$
increases with $\log{N}$, regardless of the structural classification of the proteins and the slope is higher than the
random controls. This property, along with high $C$, indicate that protein networks are ``small-world networks''~\cite{watts:book}.

\subsection{Degree Distribution}
\begin{figure*}
\begin{center}
\begin{tabular}{c}
\includegraphics[width=12cm]{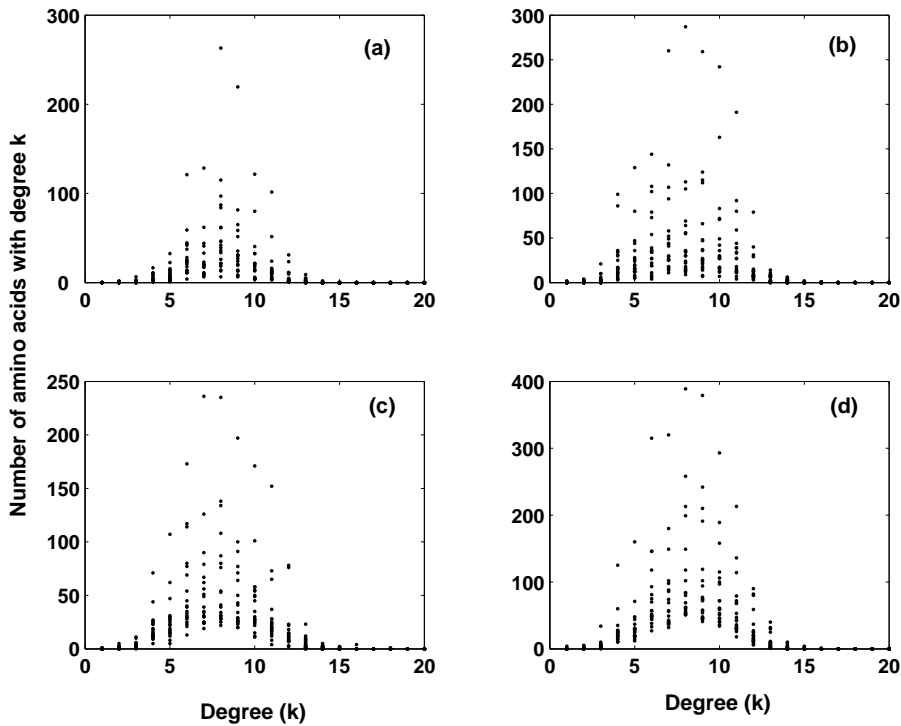}
\end{tabular}
\end{center}
\caption{Degree distributions for (a) $\alpha$, (b) $\beta$, (c) $\alpha+\beta$, and (d) $\alpha/\beta$ proteins, $20$
of each class.}
\label{fig:fig2}
\end{figure*}

The distribution of the degrees is an important property which characterises the network topology. The degree distribution of a
random network is characterised by a Poisson distribution. Figure~\ref{fig:fig2} shows the degree distributions of
$\alpha$, $\beta$, $\alpha+\beta$, and $\alpha/\beta$ protein networks. The shape of these distributions are bell-shaped,
Poisson-like~\cite{protnet:JMB}, and the number of nodes with very high degree falls off rapidly. This is understandable as there is
a physical limit on the number of amino acids that can occupy the space within a certain distance around another amino acid. Such
system-specific restrictions have been identified to be responsible for the emergence of different classes of networks
with characteristic degree-distributions by Amaral et al.\/~\cite{amaral:PNAS}. They observed that preferential
attachment to vertices in many real scale-free networks~\cite{reka:thesis} can be hindered by factors like ageing
of the vertices (e.g. actors networks), cost of adding links to the vertices, or, the limited capacity of a vertex (e.g. airports
network).

\subsection{Fibrous proteins}
\begin{figure*}
\begin{center}
\begin{tabular}{cc}
\includegraphics[width=7cm]{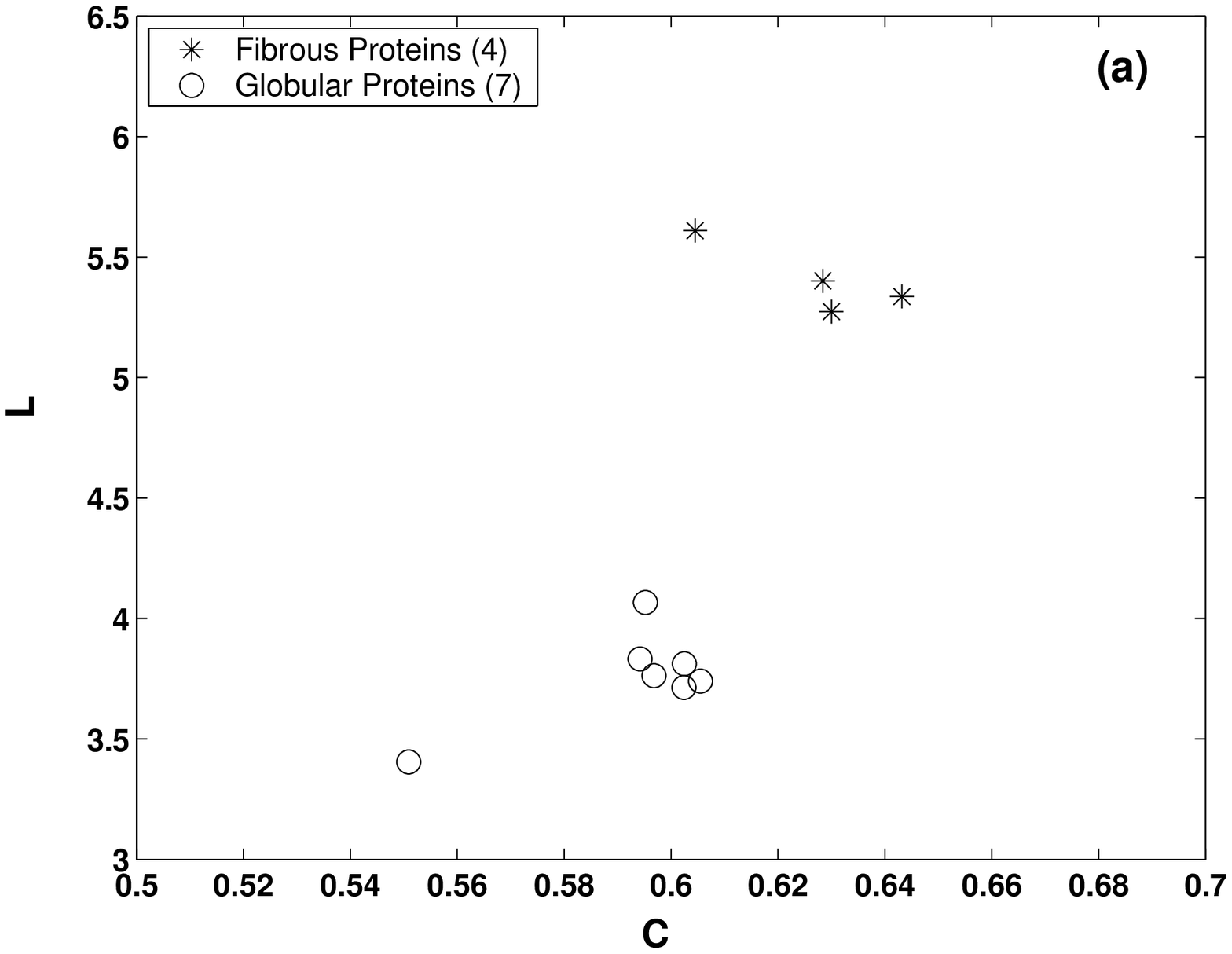} &
\includegraphics[scale=0.35]{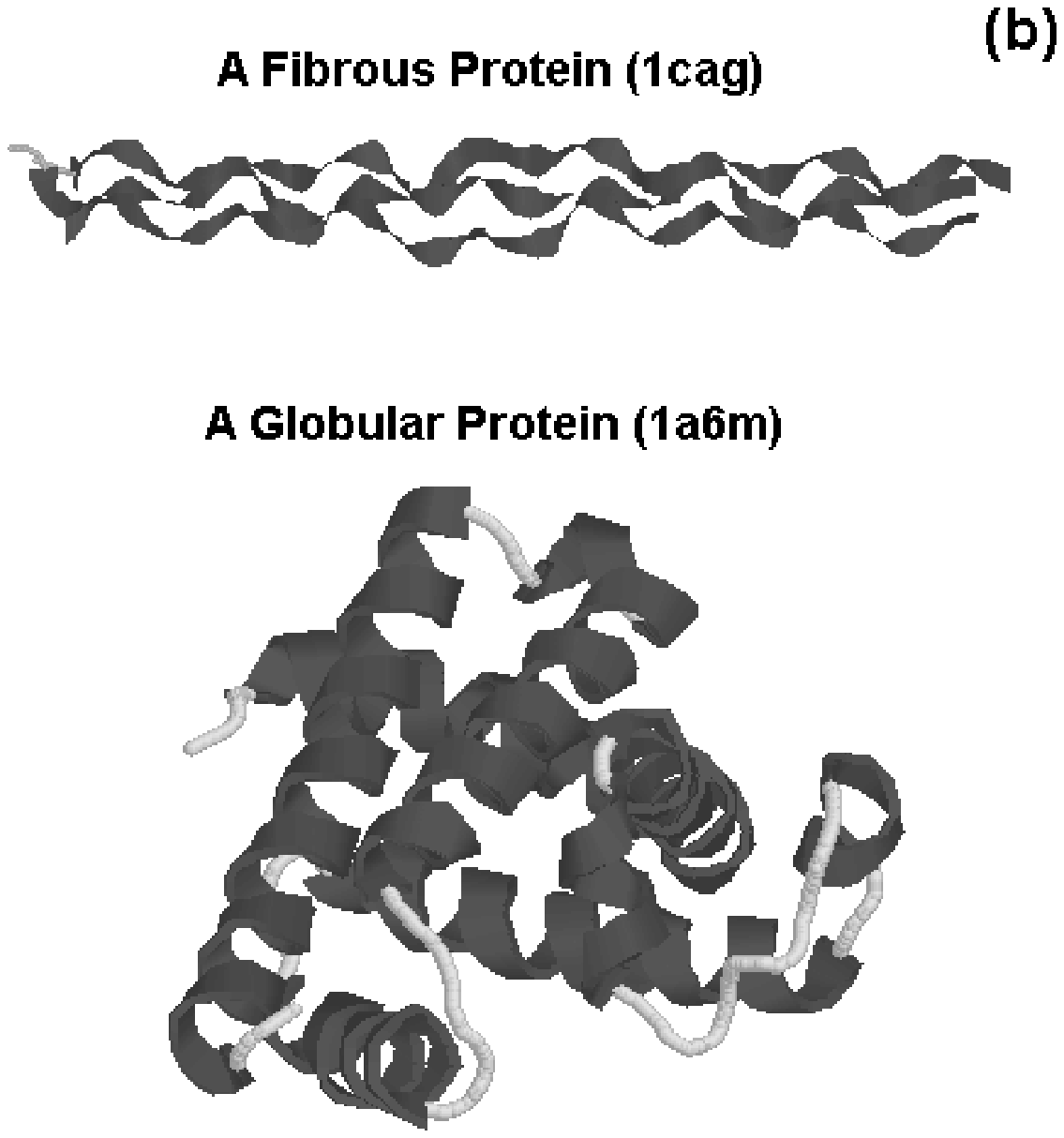}
\end{tabular}
\end{center}
\caption{(a) L--C plot of Fibrous and Globular proteins. (b) Examples of three-dimensional structures of a fibrous and globular
protein (not to the scale) with their PDB codes.}
\label{fig:fig3}
\end{figure*}

Most proteins are ``globular proteins'' in their three-dimensional structure, where the polypeptide chain folds into a compact shape.
In contrast, ``fibrous proteins'' have relatively simple, elongated three-dimensional structure suitable for their biological 
function (see Fig.~\ref{fig:fig3}(b)). 
The ``small-world'' nature of globular proteins was argued~\cite{protnet:Biophys} to be required 
for enhancing the ease of dissipation of disturbances. 
We studied fibrous proteins and compared their network properties with globular proteins of comparable size. 
As shown in the L--C plot in Fig.\/~\ref{fig:fig3}(a), fibrous proteins have larger $L$, although the $C$ are similar to those 
of globular proteins. Thus, in this respect, the fibrous proteins also show ``small-world'' properties. The average diameter for 
the fibrous 
proteins ($D=15$) was found to be larger than that of the globular proteins ($D=8.57$). This is expected because of the elongated 
structure of fibrous proteins. Despite this major difference in structure, the network properties are not much different between 
the fibrous proteins and globular proteins. This indicates that the ``small-world'' property of proteins is ubiquitous and persists 
irrespective of structural differences.

\subsection{$\alpha$ and $\beta$ proteins}
As seen earlier, both $\alpha$ and $\beta$ proteins show small-world properties. On finer analysis,
we find that there is a marginal, yet consistent difference in the $C$ of $\alpha$ and $\beta$ proteins as shown in
Fig.~\ref{fig:fig4}(a). The mean of $C$ for $\alpha$ and $\beta$ proteins studied are $0.588$ and $0.538$, respectively.
According to Kolmogorov-Smirnov test, this difference is statistically significant. Owing to the helical structure of the
$\alpha$ proteins, the amino acids are densely packed compared to that of the flat $\beta$ sheets. This may contribute to the small
increase in the Average Clustering Coefficient of the $\alpha$ proteins. Since $\alpha+\beta$ and $\alpha/\beta$ have a mixed
composition of $\alpha$ helices and $\beta$ sheets, they do not show any clear distinction.

\subsection{Change in $C$ with $N$}
\begin{figure*}
\begin{center}
\begin{tabular}{cc}
\includegraphics[width=6.5cm]{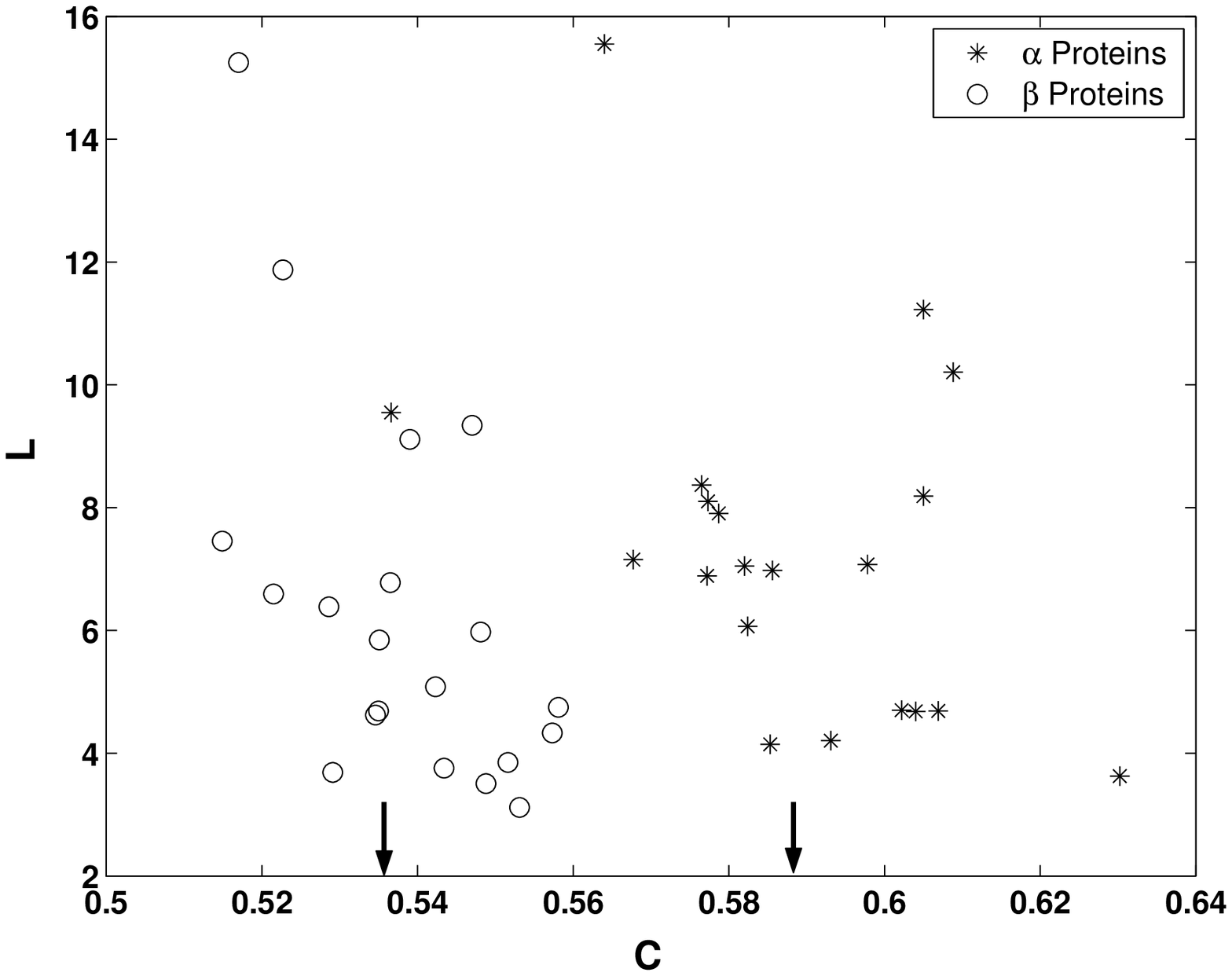} &
\includegraphics[width=6.8cm]{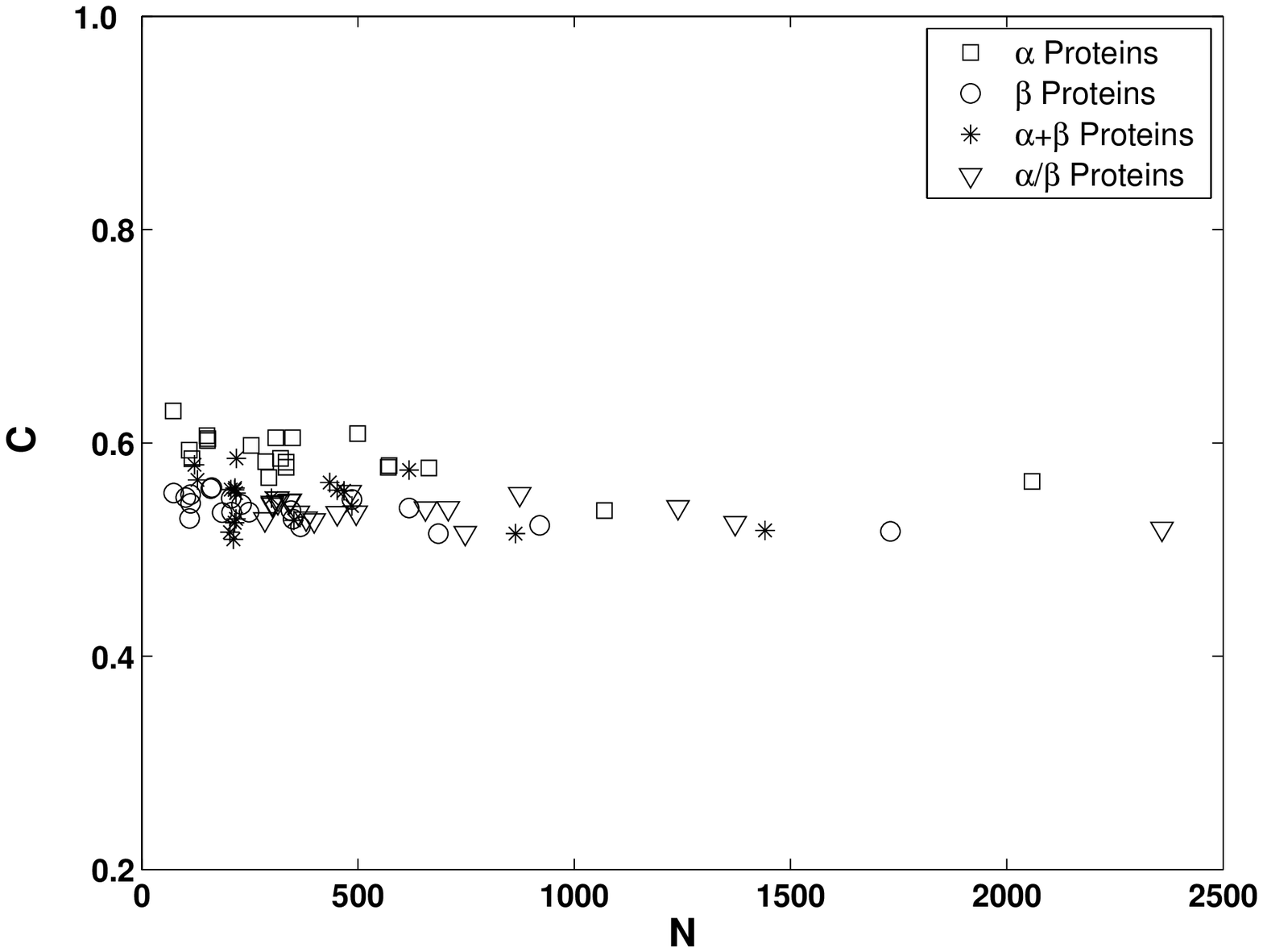}
\end{tabular}
\end{center}
\caption{(a) L--C plot for $\alpha$ and $\beta$ proteins. Arrows indicate the means of $C$ for
$\alpha$ and $\beta$ proteins.(b) Change in $C$ of proteins with increasing $N$.} 
\label{fig:fig4}
\end{figure*}

Clustering coefficient characterises local organisation. For both random as well as ``scale-free'' networks, the $C$ is expected to
fall with increasing size~\cite{reka:thesis}. It has been shown~\cite{protnet:Biophys} that, regardless of the size, the $C$
remains almost same in the core of the protein. We show the change in $C$ with increasing protein size ($N$) in
Fig.\/~\ref{fig:fig4}(b). The figure shows that $C$ does not change significantly with increasing size of proteins. Similar
result~\cite{ravsaz:science} is shown for the metabolic networks of $43$ distinct organisms. This property is suggestive of
potential modularity in the topology of the protein networks.

\section{Conclusions}
\label{sec:conclusion} 
Our results show that protein networks have ``small-world'' property regardless of their structural classification 
($\alpha$, $\beta$, $\alpha+\beta$, and $\alpha/\beta$) and tertiary structures (globular and fibrous proteins), even though small 
but definite differences exist between $\alpha$ and $\beta$ classes, and fibrous and globular proteins. The size independence of the 
Average Clustering Coefficient in proteins indicates toward an inherent modular organisation in the protein network.

In the cell, starting from a linear chain of amino acids, the protein folds in different secondary motifs such as, the $\alpha$
helices and $\beta$ sheets and their mixtures. These then assume three-dimensional tertiary structures with helices, sheets and
random coils, folding to give the final shape that is useful to carry on the biochemical function. This structure evolves in such
a way as to confer stability and also allow transmission of biochemical activity (binding of ligand, allostery, etc.\/) for
efficient functioning. Thus the networks built from such proteins are expected to show high clustering and also reflect its modular
or hierarchically folded organisation. Unlike other hierarchical networks~\cite{ravsaz:science} that are modelled to form by
replicating a core set of nodes and links, this network primarily grows linearly first, and then this polypeptide chain organises
itself in a modular manner at different levels (secondary and tertiary). Evolution of such type of network architectures demands
further study.

\begin{acknowledgments}
GB thanks Council of Scientific and Industrial Research (CSIR), Govt.\ of India, for the Senior Research Fellowship. The authors
thank Bosiljka Tadic for discussions.
\end{acknowledgments}

\end{document}